\documentclass[11pt,a4paper]{article}
\pdfoutput=1

\usepackage{jcappub}
\usepackage[utf8]{inputenc}
\usepackage{graphicx}
\usepackage{epsfig}
\usepackage{subfigure}
\usepackage{color}
\usepackage{comment}
\usepackage{amsmath}
\usepackage{mathtools}
\usepackage{tabularx}

\newcommand{\be}{\begin{equation}} 
\newcommand{\ee}{\end{equation}}
\newcommand{\bea}{\begin{equation}\begin{aligned}} 
\newcommand{\eea}{\end{aligned}\end{equation}}

\newcommand{\bmp}{\noindent\begin{minipage}{16cm}}
\newcommand{\emp}{\end{minipage}\vskip 7mm} 
\def\lsim{\mathrel{\raise.3ex\hbox{$<$\kern-.75em\lower1ex\hbox{$\sim$}}}}
\def\gsim{\mathrel{\raise.3ex\hbox{$>$\kern-.75em\lower1ex\hbox{$\sim$}}}}

\newcommand{\intron}[1]{}

\providecommand{\f}[2]{\frac{{#1}}{{#2}}}

\title{Scalar correlation functions for a double-well potential in de Sitter space}

\author[a,b]{Tommi Markkanen}
\author[c]{and Arttu Rajantie}

\affiliation[a]{Laboratory of High Energy and Computational Physics, National Institute of Chemical Physics and
Biophysics, R\"avala pst. 10, Tallinn, 10143, Estonia}
\affiliation[b]{Helsinki Institute of Physics, P.O. Box 64, FIN-00014 University of Helsinki, Finland}
\affiliation[c]{Department of Physics, Imperial College London, London, SW7 2AZ, United Kingdom}

\emailAdd{tommi.markkanen@kbfi.ee}
\emailAdd{a.rajantie@imperial.ac.uk}

\abstract{
{
We use {the spectral representation of }the stochastic Starobinsky-Yokoyama approach to compute correlation functions in de Sitter space for a scalar field with a symmetric or asymmetric double-well potential. The terms in the spectral expansion are determined by the eigenvalues and eigenfunctions of the time-independent Fokker-Planck differential operator, and we solve them numerically. The long-distance asymptotic behaviour is given by the lowest state in the spectrum, but we demonstrate that the magnitude of the coeffients of different terms can be very different, and the correlator can be dominated by different terms at different distances. This can give rise to potentially observable cosmological signatures. In many cases the dominant states in the expansion do not correspond to small fluctuations around a minimum of the potential and are therefore not visible in perturbation theory. We discuss the physical interpretation these states, which can be present even when the potential has only one minimum.}
}

\begin{document}
\begin{flushleft}
	\hfill		  IMPERIAL-TP-2020-AR-1 \\
\end{flushleft}
\maketitle
\section{Introduction}
The study of a quantized scalar field in de Sitter space is a mature endeavour \cite{Chernikov:1968zm,Dowker:1975tf,Bunch:1978yq,Birrell:1982ix}
possessing well-known difficulties when the field is light \cite{Linde:1982uu,Allen:1985ux,Allen:1987tz,Gorbenko:2019rza}. Light spectator scalars in de Sitter space are not just of formal interest, but rather they can have a variety of cosmological implications such as generation of dark matter \cite{Peebles:1999fz,Hu:2000ke,Markkanen:2018gcw} or triggering electroweak vacuum decay~\cite{Espinosa:2007qp,Herranen:2014cua,Markkanen:2018pdo}. The stochastic approach presented in Refs.~\cite{Starobinsky:1986fx,Starobinsky:1994bd} is a powerful way of addressing this problem; it is analytically tractable yet it provides accurate results that are often superior to more traditional resummation methods. The approach is based on the realization that the ultraviolet part of the field may to a good approximation be treated as white noise allowing one to express all results via classical stochastics. For other techniques, see
Refs.~\cite{Hu:1986cv,Boyanovsky:2005sh,Serreau:2011fu,Herranen:2013raa,Gautier:2013aoa,Gautier:2015pca,Tokuda:2017fdh,Arai:2011dd,Guilleux:2015pma,Prokopec:2017vxx,Moreau:2018ena,Moreau:2018lmz,LopezNacir:2019ord}

The stochastic approach has become increasingly popular in recent years, likely due to its great efficacy, and in this vein we note the recent  works~\cite{Rigopoulos:2016oko,Tokuda:2018eqs,Cruces:2018cvq,Glavan:2017jye,Hardwick:2017fjo,Vennin:2015hra,Moss:2016uix,Grain:2017dqa,Firouzjahi:2018vet,Pinol:2018euk,Hardwick:2019uex,Fumagalli:2019ohr,Jain:2019wxo,Moreau:2019jpn,Pattison:2019hef,Prokopec:2019srf,Franciolini:2018ebs}. Often the focus is on 
{the local probability distribution of the field or on local expectation values,
even though} the correlation of fluctutations over space is arguably the more relevant object physically. The spatial correlators have been addressed for example in Refs.~\cite{Peebles:1999fz,Tsamis:2005hd,Prokopec:2007ak,Riotto:2008mv,Kunimitsu:2012xx,Motohashi:2012bb,Garbrecht:2013coa,Garbrecht:2014dca,Burgess:2014eoa,Onemli:2015pma,Prokopec:2015owa,Cho:2015pwa,Vennin:2015hra,Kitamoto:2018dek,Markkanen:2018gcw,Markkanen:2019kpv}. Specifically, in Ref.~\cite{Markkanen:2019kpv} it was shown how correlators at noncoincident points can be effectively calculated with numerical techniques in conjunction with the spectral expansion based on eigenfunctions and -values already discussed in Ref.~\cite{Starobinsky:1986fx}.

The focus of Ref.~\cite{Markkanen:2019kpv} was on a single spectator scalar $\phi$ with quadratic and quartic terms in its potential while limiting the parameters to only include positive mass terms i.e. potentials possessing a single minimum at the origin. In this work we extend this analysis to include potentials with two possibly non-degenerate minima. Namely, we  focus on a potential of the form
\begin{equation}
V(\phi)=\mu^3\phi+\frac{1}{2}m^2\phi^2+\frac{\lambda}{4}\phi^4\,,\label{eq:pot}
\end{equation}
with $m^2\leq0$ and $\mu\geq0$. Throughout we will make use of the parametrization\footnote{This is related to the definition in Ref. \cite{Markkanen:2019kpv} as $\bar{\alpha}\equiv-\alpha$.}
\be
\bar{\alpha}\equiv \f{-m^2}{\sqrt{\lambda} H^2} \,;\quad \beta\equiv\f{\mu^3}{ \lambda^{1/4} H^3}\,;\quad\bar{m}^2\equiv-m^2 \,.\label{eq:ab}
\ee
{This potential has two minima if {$\bar{\alpha}>3(\beta/2)^{2/3}$}.
In typical perturbative treatments, the field is assumed to fluctuate near the minimum of the potential, but the stochastic spectral expansion does not require that assumption. As we will show, in many cases the dominant contributions to the correlators come from large-amplitude fluctuations that are not visible in perturbation theory.
}

{This paper is organised as follows: In Section~\ref{sec:spectral} we summarise the stochastic spectral expansion and the numerical technique we used to find the terms in the expansion. In Sections~\ref{sec:DBs} and \ref{sec:DBa}, we apply these techniques to calculate the spectral expansion in the symmetric and asymmetric cases, respectively. In Section~\ref{sec:Discussion} we discuss the physical interpretation and implications of our results, and in Section~\ref{sec:Conclusions} we summarise our conclusions.}

\section{The stochastic spectral expansion}
\label{sec:spectral}
\subsection{The eigenvalue equation}
%
In the stochastic formalism one may express {any correlator} involving noncoincident spacetime points {as a spectral expansion \cite{Starobinsky:1994bd,Markkanen:2019kpv}, by solving the} 
eigenvalues and eigenfunctions, $\Lambda_n$ and $\psi_n$, 
from {the eigenvalue equation}
\begin{equation}
{D}_\phi\psi_n
=-\frac{4\pi^2\Lambda_n}{H^3}\psi_n\,,\label{e:sch}
\end{equation} 
with
\begin{equation}
{D}_\phi=\frac{1}{2}\frac{\partial^2}{\partial\phi^2}-\frac{1}{2}W(\phi)\,;
\quad
v(\phi)=\frac{4\pi^2}{3H^4}V(\phi)\,;\quad W(\phi)=v'(\phi)^2-v''(\phi)\,.\label{eq:defs}
\end{equation}
\begin{figure}[t]
\begin{center}
\includegraphics[width=0.98\textwidth,trim={0cm 0 0 0cm},clip]{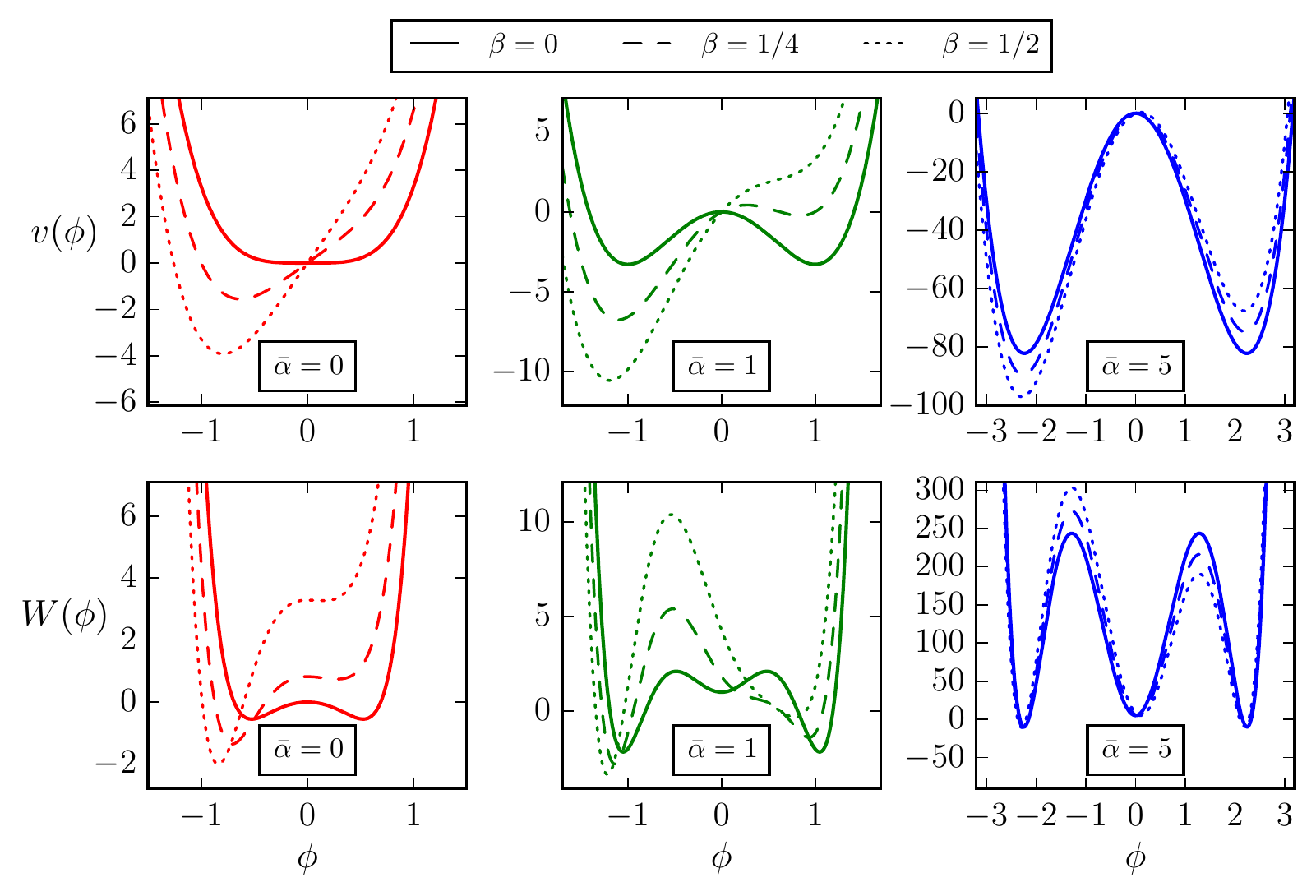}
\end{center}
\caption{\label{fig:po1}
The $v(\phi)$ and $W(\phi)$ potentials as defined in (\ref{eq:defs}) with the choice $\lambda=1$ given in the units $H=1$.}
\end{figure}Since the input in the eigenvalue equation (\ref{e:sch}) is $W(\phi)$ (and not $V(\phi)$) it is the $W(\phi)$ that will turn out to be the fundamental quantity providing a qualitative understanding of the behaviour of the eigenfunctions and -values. The behavior of $v(\phi)$ and $W(\phi)$ for the potential (\ref{eq:pot}) is illustrated in Fig. \ref{fig:po1}. In most cases solving equation (\ref{e:sch}) needs numerical methods.
\subsection{Correlators and the power spectrum}
If $G_f(t_2,t_1;\mathbf{x}_2,\mathbf{x}_1)$ is the general {correlator} 
for some function of the field $f(\phi)$, the autocorrelation function
\be
G_f(t;0)
=\langle f(\phi(0)) f(\phi(t))\rangle
,
\label{equ:Gfdef}
\ee
can via the stochastic formalism be expressed as a spectral expansion \cite{Starobinsky:1994bd,Markkanen:2019kpv} making use of the eigenfunctions and -values from Eq. (\ref{e:sch}). Specifically, in terms of eigenvalues $\Lambda_n$ and spectral coefficients $f_n$ one may write
\be
G_f(t;0)=\sum_n f_n^2 e^{-\Lambda_nt}\,,
\label{equ:spectral}
\ee
where the spectral coefficients are specific to the form of $f(\phi)$ 
\begin{equation}
f_n\equiv\int d\phi \psi_0f(\phi)\psi_n\,.\label{eq:fn}
\end{equation}
By making use of the de Sitter invariance of the equilibrium quantum state, from (\ref{equ:spectral}) one may infer the form of the equal time correlator between two spatially separated points
\be
G_f(0;\mathbf{x})
=\sum_n \frac{f_n^2}{\left(|\mathbf{x}|H
\right)^{2\Lambda_n/H}}\,,
\label{equ:spectralx}
\ee
{where $\mathbf{x}$ is physical.}
\subsection{A numerical approach}
{Following} 
Ref. \cite{Markkanen:2019kpv},
{we solve the eigenfunctions and eigenvalues of Eq. (\ref{e:sch}) numerically} with the 'overshoot/undershoot' or otherwise known as 'wag the dog' method. For a potential with $\mathbb{Z}_2$ symmetry this problem reduces to a systematic iteration of just one unknown variable, the eigenvalue $\Lambda_n$. This is due to the general feature that for a symmetric potential the wave functions are either even or odd, which fixes one of the two initial conditions required for a second order differential equation and the remaining one is irrelevant as the functions must be normalized. The essentials of the method are well-known and can be found for example in Section 2.3 of Ref.~\cite{griffiths}. This special case is encoutered in the symmetric double-well potential, which is addressed in Sec. \ref{sec:DBs}. A major shortcoming is that modifications are required when two eigenvalues are (almost) degenerate, which is encountered at the limit of a large barrier, or equivalently, deep wells. However as we will show, this is precisely the limit for which one may easily derive accurate analytic approximations and with a combination of analytics an numerics the entire eigenvalue spectrum may be covered.

When there is no $\mathbb{Z}_2$ symmetry, for example as in Eq.~(\ref{eq:pot}) with $\beta\neq0$ that is discussed in  Sec. \ref{sec:DBa}, the 'overshoot/undershoot' needs to be performed with respect to two unknowns: in addition to the eigenvalue one must iterate over the value or derivative of the function at a point giving rise to a two dimensional iteration problem. This however amounts to only a small increase in complexity of the algorithm or the expensiveness of the computation.

To facilitate a numerical analysis of the eigenfunctions and -values very similarly to Ref.~\cite{Markkanen:2019kpv} for the potential (\ref{eq:pot}) it is convenient to introduce the following dimensionless quantities
\begin{align}
    z &\equiv \f{\lambda^{1/4}\Omega}{H}\phi, \quad \Omega\equiv\bigg(1+\frac{{\bar{m}}}{H\lambda^{1/4}}+\f{\mu^3}{\lambda^{1/4}H^3}\bigg)\equiv\left(1+\sqrt{\bar{\alpha}}+\beta\right),\label{eq:fsca2}\\\tilde{\Lambda}_n&\equiv\frac{\Lambda_n}{\lambda^{1/2}H +\bar{m}^2/H+\mu^6/H^5}\equiv\frac{\Lambda_n}{\lambda^{1/2} H (1+\bar{\alpha}+\beta^2)}\,,
    \label{scaledlambda2}
\end{align}
so that the eigenvalue equation (\ref{e:sch}) may be written in terms of dimensionless numbers as
\begin{align}
  \bigg\{\frac{\partial^2}{\partial z^2}-4 \pi ^2 \frac{\bar{\alpha}+\f{4\pi^2}{3}\beta^2}{3 \Omega ^2}+\frac{32 \pi ^4 \bar{\alpha}  \beta  z}{9 \Omega ^3}&+4 \pi ^2\frac{1-\frac{4\pi ^2}{9}\bar{\alpha}^2 }{\Omega ^4}z^2-\frac{32 \pi ^4 \beta  z^3}{9 \Omega ^5}+\frac{32 \pi ^4  \bar{\alpha} z^4}{9 \Omega ^6}-\frac{16 \pi ^4 z^6}{9 \Omega ^8}\nonumber \\&+\frac{8 \pi ^2 {\left(1+\bar{\alpha}+\beta^2\right)\tilde{\Lambda}_n}}{\Omega ^2}\bigg\}\psi_n=0\,,\label{eq:quarz2}
\end{align}
with the $\bar{\alpha}$ and $\beta$ given in (\ref{eq:ab}).
Furthermore and precisely as in Ref.~\cite{Markkanen:2019kpv} we also introduce scaled eigenfunctions
\begin{equation}
    \psi_n\equiv \sqrt{\frac{\lambda^{1/4}\Omega}{H}}\tilde{\psi}_n\quad\Rightarrow\quad \int^{\infty}_{-\infty} d\phi\, | \psi_n|^2=\int^{\infty}_{-\infty} dz\, | \tilde{\psi}_n|^2=1\,. \label{eq:scaledefs}
\end{equation}

There are two reasons why the  redefinitions we introduced are very useful for numerical work. First, as one can see from Eq. (\ref{eq:quarz2}), unlike the potential that is a function of $\mu$, $m$ and $\lambda$ there are only two unknown parameters, $\bar{\alpha}$ and $\beta$. Second, there are no terms in the equation that would grow without bound at any of the limiting cases $\bar{\alpha}\rightarrow0$, $\bar{\alpha}\rightarrow\infty$, $\beta\rightarrow0$ and  $\beta\rightarrow\infty$, so broadly speaking all numerical factors are of the same order throughout the parameter space of interest.

\begin{figure}\begin{center}
\includegraphics[width=1.0\textwidth,trim={0cm 0cm 0cm 5.5cm},clip]{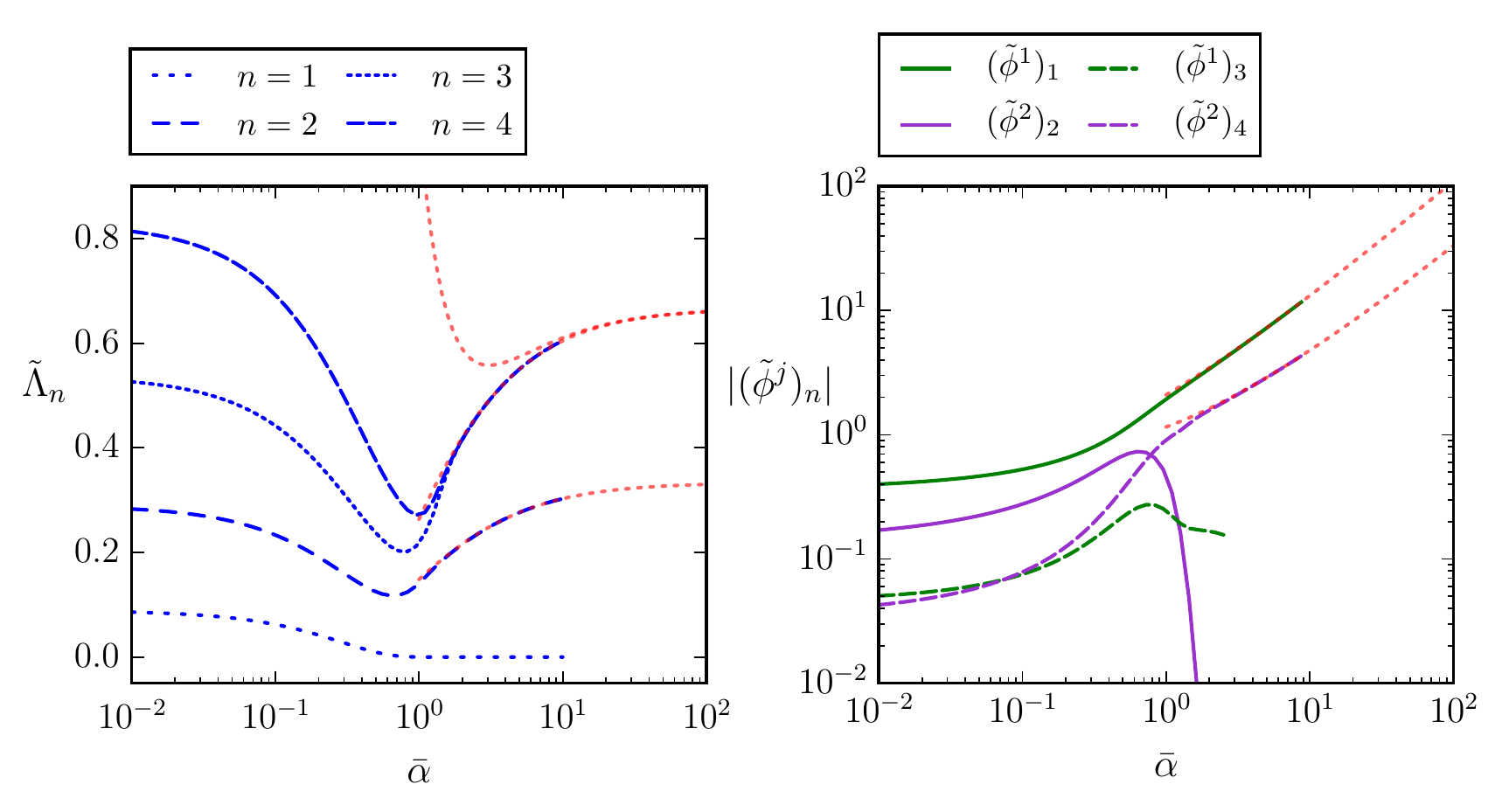}
\end{center}
\caption{\label{fig:fn}
{
The lowest eigenvalues (left) and spectral coefficients for $f(\phi)=\phi$ and $f(\phi)=\phi^2$ (right) in the symmetric case, $\beta=0$.
The red dotted lines show the analytic approximations (\ref{eq:anL2}), (\ref{eq:anL3}), (\ref{eq:anL4}), (\ref{eq:an1}) and (\ref{eq:an3}).}}
\end{figure}

\begin{figure}\begin{center}
\includegraphics[width=1\textwidth,trim={0cm 0 0 0cm},clip]{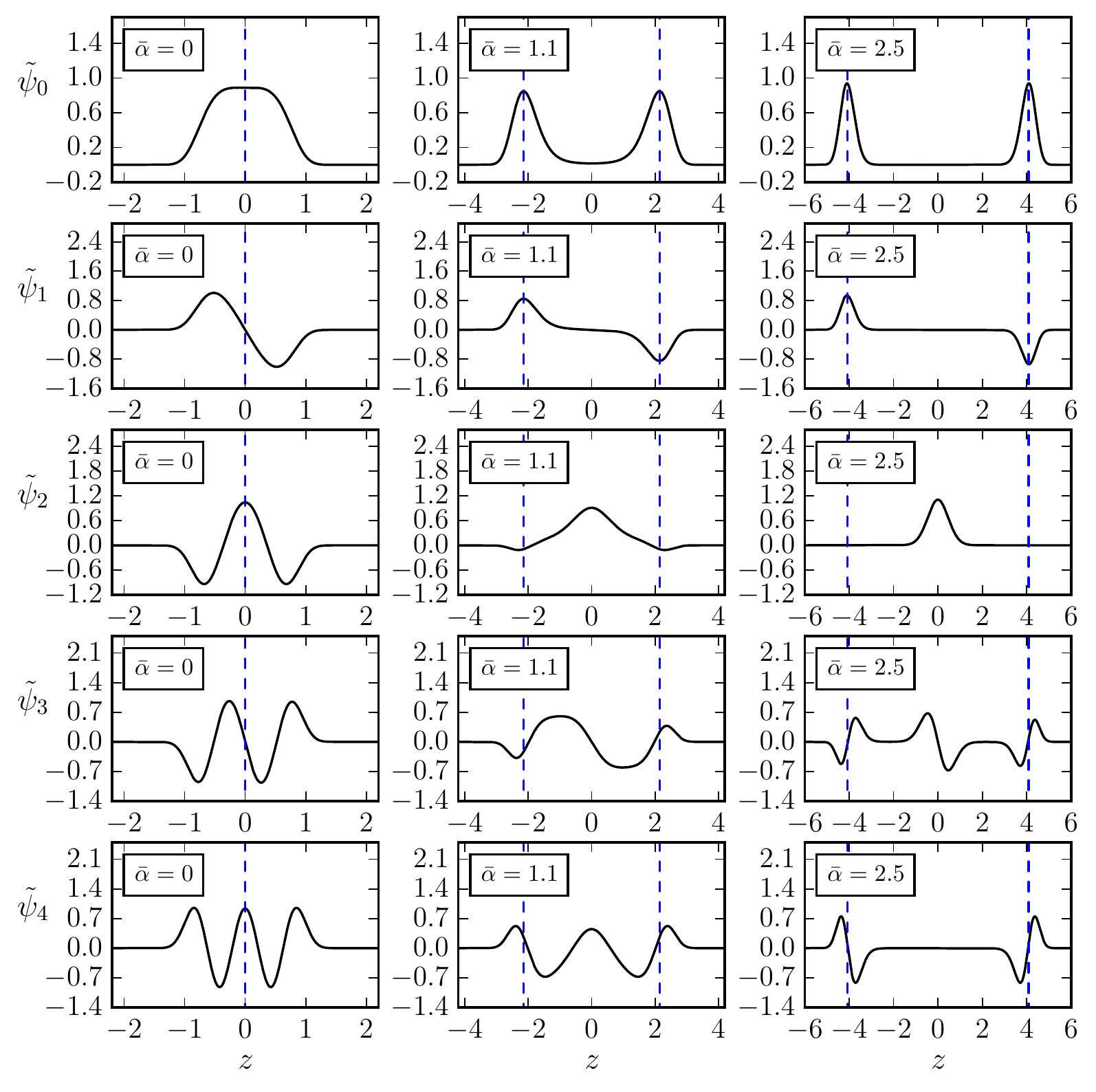}
\end{center}
\caption{
The dimensionless eigenfunctions $\tilde\psi_n$  from {Eq.}~(\ref{eq:scaledefs}) for the symmetric potential (\ref{eq:pot}) with $\beta=0$ as a function of $\bar{\alpha}\equiv \bar{m}^2/(H^2\sqrt{\lambda})$ The blue dashed lines indicate the locations of the minima of the potential $V(\phi)$.\label{fig:p}}
\end{figure}

\section{The symmetric double-well potential}
\label{sec:DBs}
\subsection{The first five eigenfunctions/-values}
\label{sec:examples}

Before addressing the general situation, {let us first focus on the important special case of a symmetric double-well potential with two degenerate minima} i.e.  with $\beta=0$. 
The first four non-trivial eigenvalues are plotted in Fig.\,\ref{fig:fn}. For $\bar{\alpha}\gtrsim 2.5$ the {eigenvalues $\Lambda_3$ and $\Lambda_4$} become degenerate, complicating the numerical analysis, but by this point analytic approximations {derived in Section~\ref{sec:dlim} (depicted with red dotted lines)} are already very accurate. 

{The eigenfunctions $\tilde\psi_n$ for $n\le 4$}
are shown in Fig.\,\ref{fig:p}. 
For $\bar{\alpha}\gtrsim1$ the system separates into linear combinations of solutions centered at the three minima of $W(\phi)$, which can be seen in Fig.\,\ref{fig:po1}. {Importantly, there are} 
solutions that do not vanish close to the origin, even when $\bar{\alpha}\gg1$ which corresponds to a double-well potential with a very large barrier. 
{The clearest example of this is $\tilde\psi_2$ at $\bar{\alpha}=2.5$, which is localised around the top of the potential barrier.
As discussed in Section~\ref{sec:Discussion},
these solutions can be interpreted as a contribution from transitions between the two minima, during which the field can spend a significant amount of time near the top of the barrier.  {The existence of such solutions was apparently not noticed in Ref.~\cite{Starobinsky:1994bd}.}
}

\subsection{The large barrier limit}
\label{sec:dlim}
As {can be seen} from Fig.\,\ref{fig:po1}, in the limit a large $\bar{\alpha}$ the potential $W(\phi)$ in the eigenvalue equation (\ref{e:sch}) will possess three minima separated by large barriers, which can be shown to occur at
\be
z_0=0\,;\qquad z_\pm=\pm\frac{\sqrt{\bar{\alpha}}+1}{\sqrt{3}} \left(\sqrt{\bar{\alpha}^2+\frac{27}{4 \pi ^2}}+2 \bar{\alpha} \right)^{1/2}\,.
\ee
Hence, at the limit $\bar{\alpha}\rightarrow\infty$, the system separates into three quadratic pieces, which can be obtained from (\ref{eq:quarz2}) by expanding around a large $\bar{\alpha}$ with $\beta=0$.

For completeness\footnote{See Ref.~\cite{Markkanen:2019kpv} for more discussion and plots.}  we first write the eigenvalue equation for $V(\phi)=\f12 M^2\phi^2$
\be
\bigg\{\frac{\partial^2}{\partial x^2}-\left(\frac{4\pi^2}{3}\right)^2 {x}^2+\frac{4\pi^2}{3}+{8 \pi ^2 {\tilde{\Lambda}_n}}\bigg\}\psi_n=0\,;\quad x=\f{M}{H^2}\phi\,;\quad\tilde{\Lambda}_n=\f{\Lambda_n}{M^2/H}\,,
\ee
with the eigenfunctions and -values

\begin{equation}
\psi_n=\frac{\sqrt{M}}{H}\frac{1}{\sqrt{2^n n!}}\bigg(\frac{4 \pi }{3 }\bigg)^{1/4} e^{-\frac{2 \pi ^2 {x}^2}{3}}H_n\left(\frac{2 \pi  {x}}{\sqrt{3}}\right)
\,;\quad \Lambda_n=\f{n}{3}\f{M^2}{H}\,.\label{eq:fulls}
\end{equation}

Close to the origin at the limit of large barriers {i.e. taking $\bar{\alpha}\rightarrow\infty$} we then get the approximate eigenvalue equation for the potential (\ref{eq:pot})
\begin{align}
\bigg\{\frac{\partial^2}{\partial z^2}-\left(\frac{4\pi^2}{3}\right)^2 {z}^2+\frac{4\pi^2}{3}+{8 \pi ^2 \bigg({\tilde{\Lambda}_n}-\f13}\bigg)\bigg\}\psi^0_n=0\,;\quad z=\f{\bar{m}}{H^2}\phi\,;\quad\tilde{\Lambda}_n=\f{\Lambda_n}{\bar{m}^2/H}\,,
\end{align}
where the eigenfunctions and -values can be read off from the quadratic results (\ref{eq:fulls})

\begin{equation}
\psi^0_n=\frac{{{\sqrt{\bar{m}}}}}{H}\frac{1}{\sqrt{2^n n!}}\bigg(\frac{4 \pi }{3 }\bigg)^{1/4} e^{-\frac{2 \pi ^2 {z}^2}{3}}H_n\left(\frac{2 \pi  {z}}{\sqrt{3}}\right)
\,;\quad \Lambda_n=\f{n+1}{3}\f{\bar{m}^2}{H}\,.\label{eq:fulls0}
\end{equation}
At $z_\pm$ for large barriers one gets the approximate equation
\begin{align}
\bigg\{\frac{\partial^2}{\partial y^2}-\left(\frac{4\pi^2}{3}\right)^2 {y}^2+\frac{4\pi^2}{3}+8 \pi ^2 \bigg(\f{\tilde{\Lambda}_n}{2}\bigg)\bigg\}\psi^\pm_n=0\,;\quad y=\sqrt{2}\big(z- z_\pm\big)\,;\quad\tilde{\Lambda}_n=\f{\Lambda_n}{\bar{m}^2/H}\,,
\end{align}
with the eigenfunctions and -values

\begin{equation}
\psi^\pm_n=\frac{\sqrt{{\bar{m}}}}{H}\frac{1}{\sqrt{2^n n!}}\bigg(\frac{8 \pi }{3 }\bigg)^{1/4} e^{-\frac{2 \pi ^2 {y}^2}{3}}H_n\left(\frac{2 \pi  {y}}{\sqrt{3}}\right)
\,;\quad \Lambda_n=\f{2n}{3}\f{\bar{m}^2}{H}\,.\label{eq:fullspm}
\end{equation}

{By making use of the approximate solutions close to the origin and/or $z_\pm$, (\ref{eq:fulls0}) and (\ref{eq:fullspm}) respectively, and using Fig.\,\ref{fig:p} as a guide is it possible to understand qualitatively the large $\bar{\alpha}$ behaviour and often write analytic approximations for the eigenfunctions and -values.

Suppose an analytic function $\Phi_n(\phi)$ that approximates the full solution. Then, by calculating the expectation value of the eigenvalue equation (\ref{e:sch}) one gets an approximation for the eigenvalue as
\be
\int d\phi\,\Phi_n D_\phi\Phi_n
\approx-\frac{4\pi^2\Lambda_n}{H^3}\label{eq:an}\,.
\ee

For example, from Fig.\,\ref{fig:p} we see that for large $\bar{\alpha}$ the $n=1$ eigenfunction approaches an antisymmetric combination of two quadratic $n=0$ solutions centered at $z_\pm$ (\ref{eq:fullspm}). The eigenvalues of the quadratic $n=0$ solutions are zero at $z_\pm$ as given by (\ref{eq:fullspm}), which implies that the first exited state has a neglibigle eigenvalue 
at this limit. As derived in \cite{Starobinsky:1994bd}, for any large but finite $\bar{\alpha}$ the $n=1$ eigenvalue is exponentially small.}

Similarly, we see that the eigenfunction for $n=2$ approaches the quadratic $n=0$ solution at the origin, which as (\ref{eq:fulls0}) shows does not result in a vanishing eigenvalue, even at $\bar{\alpha}\gg1$. The analytic estimate for the eigenvalue is obtained by
\begin{align}
\int d\phi\,\psi^0_0{D}_\phi\psi^0_0
&\approx-\frac{4\pi^2\Lambda_2}{H^3}\label{eq:anL2}
\quad\nonumber\\\Rightarrow\quad\tilde{\Lambda}_2&\approx\frac{32 \pi ^2 \left(8 \pi ^2\bar{\alpha} ^2-9\right) \bar{\alpha} ^2+135}{768 \pi ^4\bar{\alpha} ^3 (\bar{\alpha}+1)}=\f13\bigg(1-\frac{1}{\bar{\alpha} }+\frac{1-\frac{9}{8 \pi ^2}}{\bar{\alpha} ^2}\bigg)+{\cal O}(\bar{\alpha}^{-3})\,.
\end{align}

The $n=3$ case can be seen from Fig.\,\ref{fig:p} to approach a linear combination of three quadratic $n=1$ solutions centered at the origin and at $z_\pm$. From (\ref{eq:fulls0}) and (\ref{eq:fullspm}) we see that the solutions are degenerate so we can derive an analytic approximation for the eigenvalue by simply using only the solution at $z_0$ giving
\begin{align}
\label{eq:anL3}
\int d\phi\,\psi^0_1{D}_\phi\psi^0_1
&\approx\nonumber-\frac{4\pi^2\Lambda_3}{H^3}
\\\quad\Rightarrow\quad\tilde{\Lambda}_3&\approx\frac{512 \pi ^4 \bar{\alpha} ^4-1152 \pi ^2 \bar{\alpha} ^2+945}{768 \pi ^4 \bar{\alpha} ^3 (\bar{\alpha}+1)}=\f23\bigg(1-\frac{1}{\bar{\alpha} }+\frac{1-\frac{9}{4 \pi ^2}}{\bar{\alpha} ^2}\bigg)+{\cal O}(\bar{\alpha}^{-3})\,.
\end{align} 

Finally, the $n=4$ eigenfunction clearly approaches a symmetric combination of two quadratic $n=1$ solutions centered at $z_\pm$, with then the eigenvalue at the large $\bar{\alpha}$ limit approximated by
\begin{equation}
\label{eq:anL4}
\int d\phi\,\psi^\pm_1{D}_\phi\psi^\pm_1
\approx-\frac{4\pi^2\Lambda_4}{H^3}
\quad\Rightarrow\quad\tilde{\Lambda}_4\approx \f23\bigg(1-\frac{1}{\bar{\alpha} }+ \frac{1+\frac{1341}{256 \pi ^2}}{\bar{\alpha} ^2}\bigg)+{\cal O}(\bar{\alpha}^{-3})\,.
\end{equation}
In the above for $n=4$ we have only included the leading terms as the full result is quite complicated.

The analytic approximations of this section are depicted by the red dashed curves in the left panel in Fig.\,\ref{fig:fn}. As one may see, they are in very good agreement with the full results for $\bar{\alpha}\gtrsim1$. 
\subsection{The spectral coefficients}
\label{sec:spe}
In addition to the eigenvalues, the spectral coefficients are the other ingredient needed for calculating correlators as given by Eq. (\ref{equ:spectral}) and (\ref{equ:spectralx}).
As an illustration in the following we analyse the leading and next-to-leading spectral coefficients for $f(\phi)=\phi$ and $f(\phi)=\phi^2$. A useful dimensionless definition comes via (see Eq. (\ref{eq:scaledefs}))
\begin{equation}
(\tilde{\phi}^j)_n\equiv\bigg(\f{\lambda^{1/4}\Omega}{H}\bigg)^j({\phi}^j)_n=\bigg(\f{\lambda^{1/4}\Omega}{H}\bigg)^j\int^{\infty}_{-\infty} d\phi \psi_0\phi^j  \psi_n
=\int^{\infty}_{-\infty} d z\, \tilde{\psi_0}z^j  \tilde{\psi}_n\,.\label{eq:fsa}
\end{equation}
Much like for the eigenvalues, the spectral coefficients can in some circumstances be approximated with analytic results at the large barrier limit, which may be deduced from section \ref{sec:dlim} and in particular Fig.\,\ref{fig:p}. 

{Let us focus on $f(\phi)=\phi$ first. Because it is an odd function, only odd $n$ contribute. Because} 
the ground state and the first exited state approach symmetric and antisymmetric combinations of the ground state of a harmonic oscillator located at $z_\pm$, we may approximate the $(\phi^1)_1$ coefficient as
\begin{align}
|(\phi^1)_1| &\approx |\int^{\infty}_{-\infty} d\phi \f{1}{\sqrt{2}}\left(\psi^+_0+\psi^-_0\right)\phi  \f{1}{\sqrt{2}}\left(\psi^-_0-\psi^+_0\right)|\approx |\int d\phi \phi\left(\psi^\pm_0\right)^2|
\nonumber \\ \Rightarrow\quad
|(\tilde{\phi}^1)_1| &\approx |\int dz z\big(\tilde{\psi}^\pm_0\big)^2|=\frac{\left(\sqrt{\bar{\alpha}}+1\right) \sqrt{\sqrt{4 \pi ^2 \bar{\alpha}^2+27}+4 \pi  \bar{\alpha} }}{\sqrt{6 \pi }}\,.\label{eq:an1}
\end{align}

The third excited state is for large $\bar{\alpha}$ approximately a linear combination of $\psi_1^0$ and $-\psi^+_1-\psi^-_1$. { For this case in our approximative prescription there is an ambiguity as there is no a priori way to determine the relative size between $\psi_1^0$ and $-\psi^+_1-\psi^-_1$.}

{The analytic approximations (\ref{eq:an1}) and (\ref{eq:an3}), as well as the numerical results for the leading and next-to-leading spectral coefficients are shown in Fig.\,\ref{fig:fn}\,. The $|(\tilde{\phi}^1)_3|$ term, shown by the dashed green line, is cut short by our inability to extend the numerical method to cases with almost degenerate eigenvalues, which occurs for $n=3$. However, it is clearly subleading to $|(\tilde{\phi}^1)_1|$ denoted with green that does not suffer from this issue.}

{Correspondingly, because $f(\phi)=\phi^2$ is even, only even values of $n$ contribute to its correlator.}
Since there is virtually no overlap with the $n=0$ and $n=2$ solutions at the large barrier limit, 
$|(\tilde{\phi}^2)_2|$ is expected to vanish up to exponentially small terms.
{In contrast,} $(\phi^2)_4$ has the approximation
\begin{align}
|(\phi^2)_4| &\approx |\int^{\infty}_{-\infty} d\phi\f{1}{\sqrt{2}}\left(\psi^+_0+\psi^-_0\right)\phi^2 \f{1}{\sqrt{2}}\left(\psi^+_1-\psi^-_1\right)|  \approx |\int d\phi \phi^2\psi^\pm_0\psi^\pm_1|
\nonumber \\ \Rightarrow\quad
|(\tilde{\phi}^2)_4| &\approx |\int dz z^2\tilde{\psi}^\pm_0\tilde{\psi}^\pm_1|=\frac{\left(\sqrt{\bar{\alpha}}+1\right)^2 \sqrt{8 \pi +\frac{2 \sqrt{4 \pi ^2 \bar{\alpha}^2+27}}{\bar{\alpha}}}}{4 \pi ^{3/2}}\,.\label{eq:an3}
\end{align}
{This, together with the numerical results for $|(\tilde{\phi}^2)_2|$ and $|(\tilde{\phi}^2)_4|$ are shown in Fig.\,\ref{fig:fn}\,.} {The fact that $|({\phi}^2)_4|$ can dominate over $|({\phi}^2)_2|$ as seen in Fig.\,\ref{fig:fn} has important physical consequences, which are discussed in Section~\ref{sec:Discussion}.}
\begin{figure}[t]\begin{center}
\includegraphics[width=1.0\textwidth,trim={0cm 0cm 0cm 5.5cm},clip]{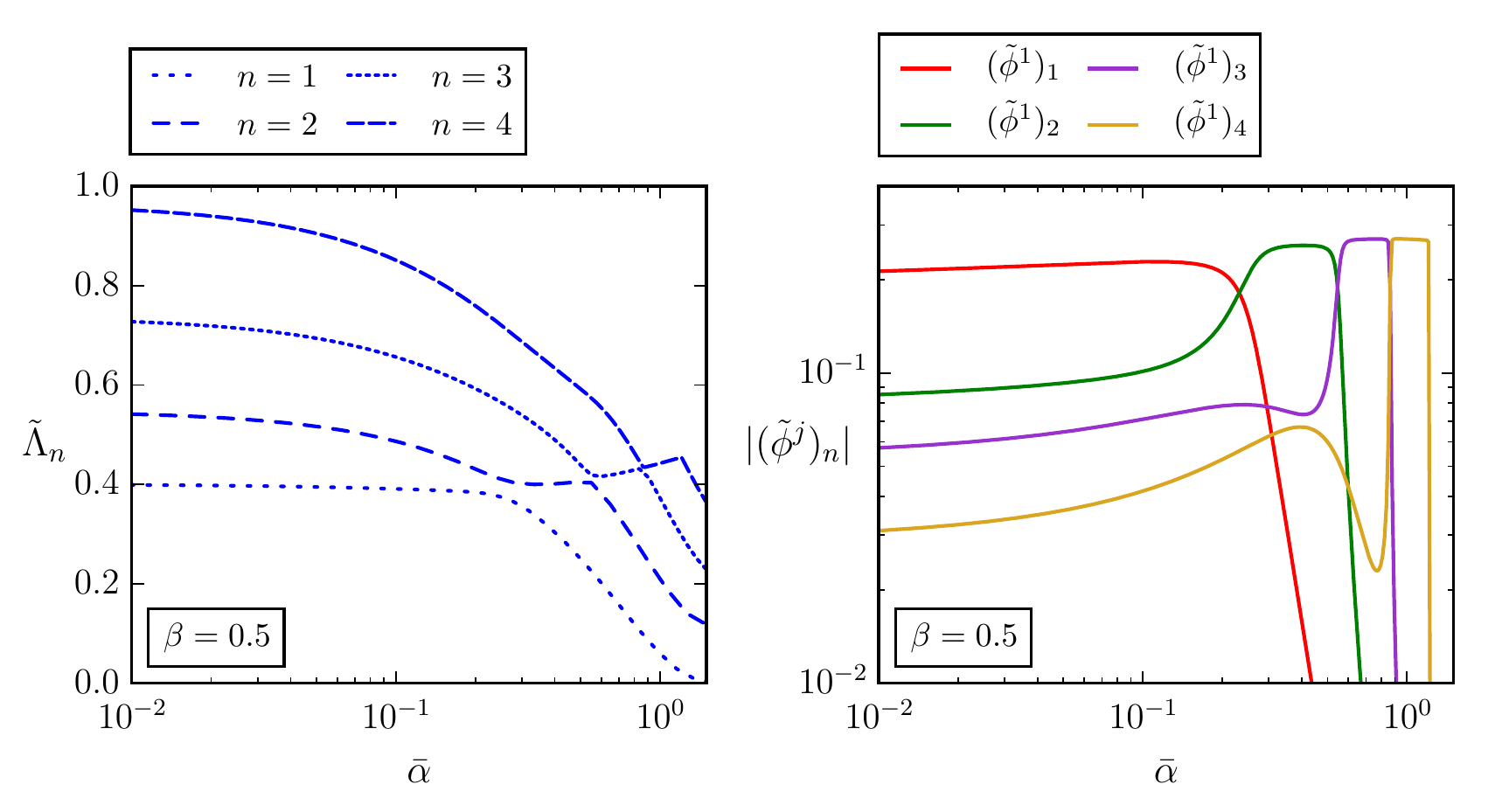}
\end{center}
\caption{\label{fig:fn0}
The eigenvalues $\tilde\Lambda_n$ (left) and the spectral coefficients for $f(\phi)=\phi$ (right) {in the asymmetric case with} $\beta=0.5$, as functions of $\bar{\alpha}$. \label{fig:b5}}
\end{figure}
\section{The asymmetric double-well potential}
\label{sec:DBa}

\begin{figure}[t]\begin{center}
\includegraphics[width=1\textwidth,trim={0cm 0cm 0cm 0cm},clip]{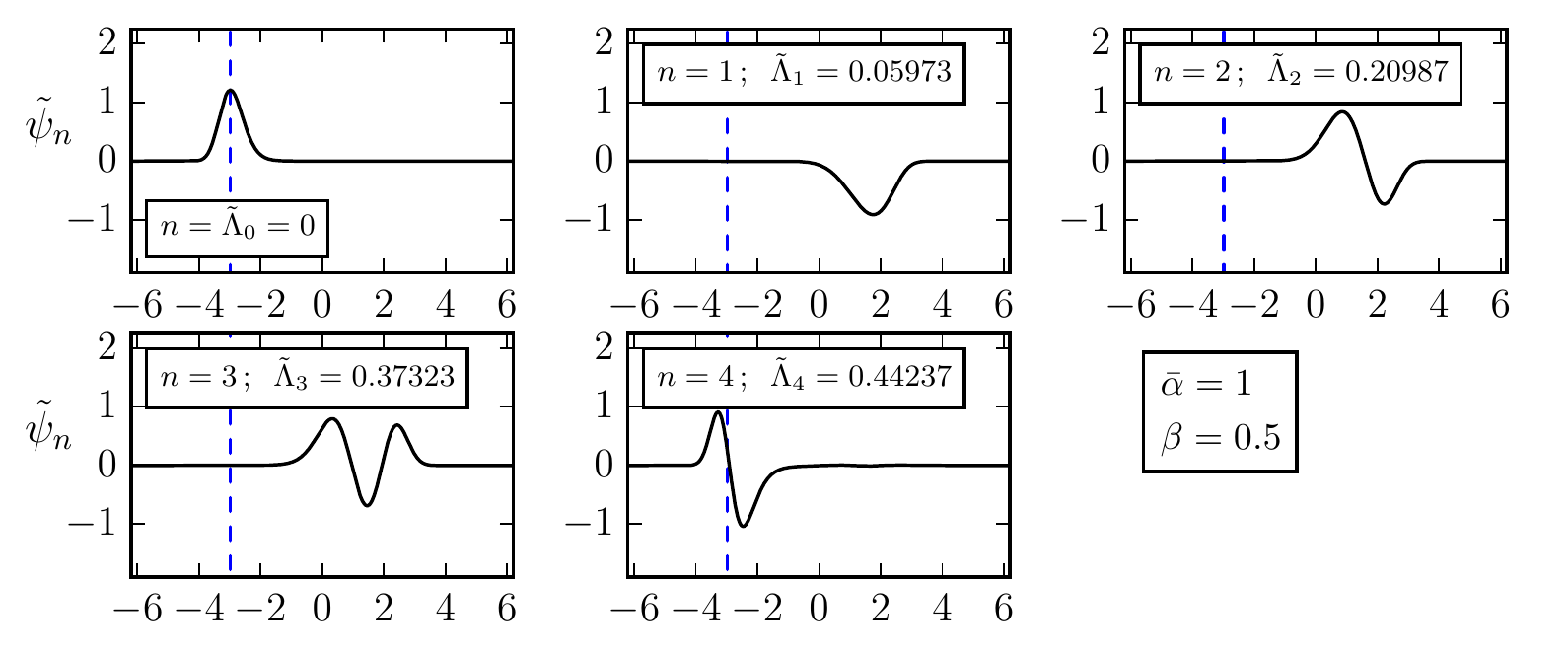}
\end{center}
\caption{\label{fig:ef}
The $n\leq4$ eigenfunctions and their    respective eigenvalues from (\ref{eq:quarz2}) with $\bar{\alpha}=1$ and $\beta=0.5$\label{fig:b6}. The blue dashed line indicates the location of the minimum of the potential.}
\end{figure}

Let us now move to the more general asymmetric case, with $\beta>0$.
Now, the potential $V(\phi)$  has a single minimum if $\bar{\alpha}<3(\beta/2)^{2/3}$, and two non-degenerate minima if $\bar{\alpha}>3(\beta/2)^{2/3}$.
On the hand, the function $W(\phi)$, which appears in the eigenvalue equation (\ref{e:sch}), seems to always have several minima.
Depending on its specific form, some of the lowest eigenfunctions may be localised at the other minima of $W(\phi)$, rather than the vacuum state located at the minimum of the potential $V(\phi)$. 
These contributions would not be visible in typical perturbation theory calculations.
They can also give rise to non-trivial hierarchies between the spectral coefficients (\ref{eq:fn}).

As an illustrative example, Fig.~\ref{fig:fn0} shows the lowest eigenvalues for $\beta=0.5$ as well as the corresponding spectral coeffients for $f(\phi)=\phi$. 
At $\bar{\alpha}=0$, the lowest eigenvalue $\Lambda_1$ is localised around the vacuum state {and therefore it corresponds to perturbative fluctuations}. However, when $\bar{\alpha}$ becomes larger, it is overtaken by other eigenvalues which are localised around the other minima of $W(\phi)$. 

{This can be seen in Fig.~\ref{fig:ef}, which shows the five lowest eigenfunctions for $\bar{\alpha}=1$ and $\beta=0.5$. Even though the potential $V(\phi)$ actually has only one minimum for these parameters (see Fig.~\ref{fig:po1}), the lowest excited state localised at the minimum of the potential is $n=4$.}
The asymptotic form of the correlator is therefore determined by the shape of the potential away from its minimum. However, because such states have a very small overlap with the ground state $\psi_0$, their spectral coefficients are very small. This is discussed more in Section~\ref{sec:Discussion}.
%
The eigenfunctions and -values up to $n=4$ for some representative choices for $\bar{\alpha}$ and $\beta$ are shown in Figs.\,\ref{fig:p1}--\ref{fig:p5} in Appendix~\ref{sec:appendix}.

\section{Discussion}
\label{sec:Discussion}


\subsection{Length scales}
In cosmology, the observable length scales correspond to comoving distances that were many orders of magnitude longer than the Hubble length during inflation. We are therefore often interested in the correlator at distances that are very long but still finite.

The asymptotic long-distance behaviour of the correlator $G_f(0,r)$ is given by the first term in the spectral expansion (\ref{equ:spectralx}) with a non-zero spectral coefficient $f_n$.
The behaviour is simplest
if  the lowest state, {$n=1$}, has the largest spectral coefficient, because then the correlator is well approximated by a single power-law at all length scales,
\be
G_f(0,r)\approx \frac{f_1^2}{(rH)^{2\Lambda_1/H}}.
\ee
However, this is not always the case, and more generally, the correlator can be dominated by a higher term $n=d$ in the expansion at the distances of interest, 
\be
G_f(0,r)\approx \frac{f_d^2}{(rH)^{2\Lambda_d/H}}.
\ee

In particular, the short-distance behaviour of the correlator can be very different from its asymptotic long-distance form.
To characterise that, {following Ref. \cite{Starobinsky:1994bd}} we define a correlation radius to be the distance where the correlator has fallen to half of its value at $r=1/H$,
\be
G_f(0,R_f)=\f12 G_f(0,1/H)
\,.
\ee
{When a single coefficient $n=d$ dominates the correlator, this simply gives}
\be
R_f\approx H^{-1}2^{\f{H}{2\Lambda_d}}\,.
\ee
The non-trivial hierarchies between different terms in the spectral expansion can be important for cosmological observations. If there is a change in the behaviour of the correlator within the observable scales, it can potentially be detected providing useful information about the fields responsible for it.

\subsection{Symmetric potential}
In the symmetric case ($\beta=0$) discussed in Section~\ref{sec:DBs}, there are two examples of this non-trivial behaviour. 
The first is that, because the lowest eigenstate $n=1$ has odd parity, the corresponding spectral coefficient vanishes for all even functions. The asymptotic long-distance behaviour of any even correlator, such as those of $\phi^2$ or the energy density, is therefore given by the second-lowest eigenvalue $\Lambda_2$. When $\bar{\alpha}\gtrsim 1$, the difference between them can be large, as we can see from the left panel in Fig.~\ref{fig:fn}. Because $\Lambda_1$ is small, the field itself is correlated over massively superhorizon scales, but its energy density is not, because its correlations are determined by $\Lambda_2$. The physical reason for this behaviour is that on superhorizon scales, the system consists of domains of the two vacua, which contribute to the field correlator, and the correlation radius $R_\phi$ gives the typical size of these domains. However, because both vacua have the same energy density, these domains do not give any contribution to correlators of even quantities such as the energy density.

The second example is that, as we can see from the right panel of Fig.~\ref{fig:fn}, the spectral coefficient of $\phi^2$ for the second-lowest state $n=2$ falls rapidly when $\bar{\alpha}\gtrsim 1$.
This means that although the asymptotic long-distance behaviour of any even correlator is indeed given by $\Lambda_2$, it only starts to dominate at very long superhorizon distances, and at shorter distances the dominant contribution is given by $\Lambda_4$. To understand why this happens, it is useful to look at the corresponding eigenfunctions in Fig.~\ref{fig:p}. The higher state $\psi_4$ is localised at the minima of the potential, and therefore it corresponds to small-amplitude perturbative fluctuations around either vacuum state. The lower state $\psi_2$, on the other hand, is localised on top of the barrier, $\phi=0$, which shows that this contribution comes from the boundaries between the domains. The eigenvalue $\Lambda_2$ characterises the thickness of these domain walls, and the spectral coefficient is small because their volume is small compared with the volume of the domains.

\begin{figure}[t]\begin{center}
\includegraphics[width=1.0\textwidth,trim={0cm 0cm 0cm 4.3cm},clip]{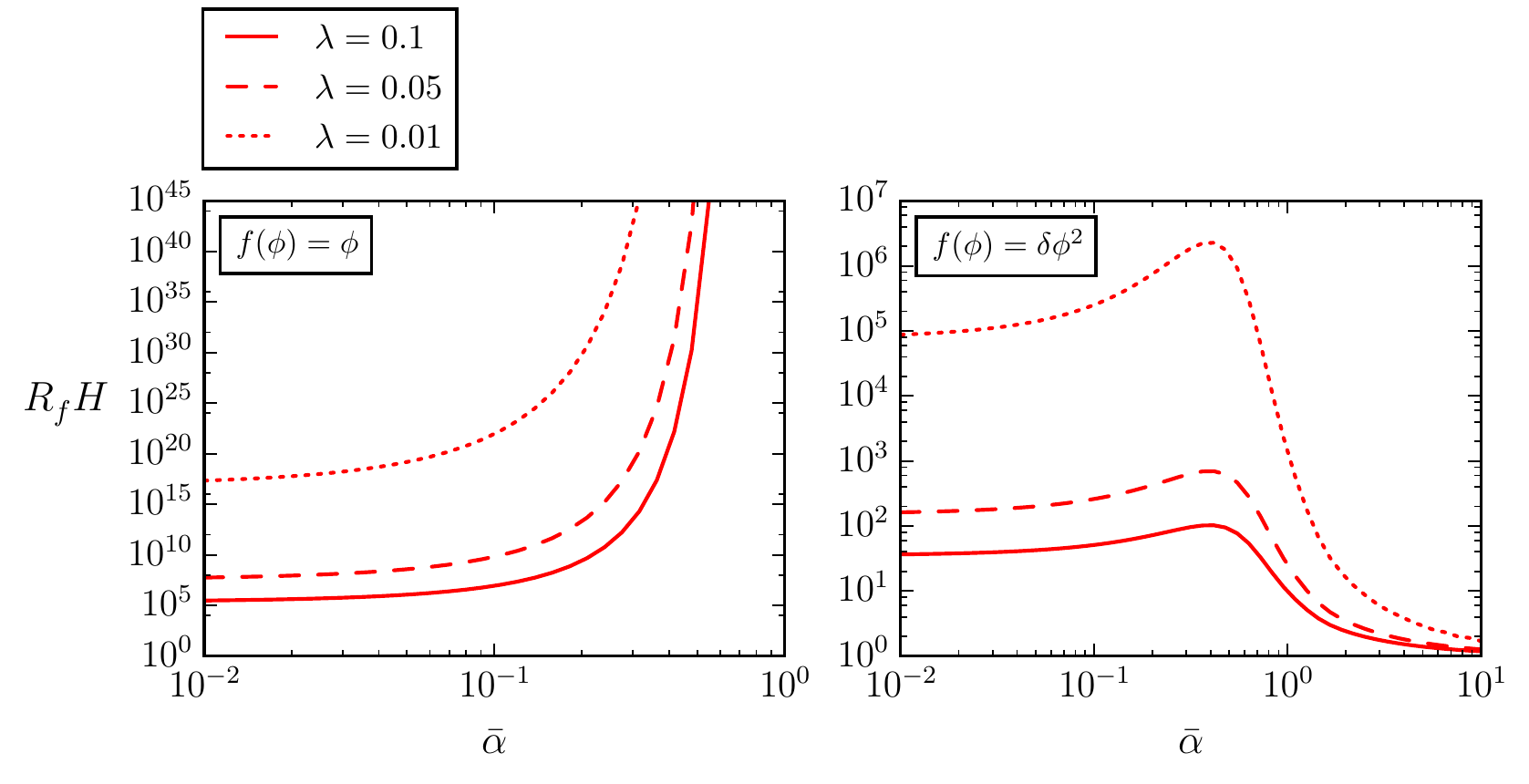}
\end{center}
\caption{The correlation radii for $f( \phi)=\phi$ and $f( \phi)=\delta\phi^2\equiv\phi^2-\langle\phi^2\rangle$ for three values of $\lambda$. \label{fig:R}
}
\end{figure}

The different behaviour of odd and even correlators is illustrated by Fig.~\ref{fig:R}, which shows the correlation radii of $\phi$ and $\delta\phi^2$. As $\Lambda_1$ becomes small, the field correlation radius grows to massively superhorizon scales. In contrast, when $\bar{\alpha}\gtrsim 1$, the correlation radius of $\delta\phi^2$ starts to decrease, first because $\Lambda_2$ grows, and then because $\Lambda_4$ starts to dominate it.

If any variable has significant correlations at distances much longer than the current Hubble length $1/H_0$, they would appear as a homogeneous background value. In the case of a symmetric potential, this happens when the domain size is larger than $1/H_0$, in which case the current observable Universe would most likely be inside single domain. In that case the observed field correlator would be determined by the second-lowest odd eigenvalue $\Lambda_3$.

\subsection{Asymmetric potential}

In the asymmetric case discussed in Section~\ref{sec:DBa}, we can see further examples of these non-trivial hierarchies. When $\bar{\alpha}$ is small, the lowest state $n=1$ corresponds to perturbative fluctuations around the true vacuum. However, as $\bar{\alpha}$ increases, other states overtake it one by one, as we can see from Fig.~\ref{fig:b5}. 

Fig.~\ref{fig:b6} shows the eigenfunctions for $\bar{\alpha}=1$ and $\beta=0.5$, and illustrates that for these parameters the lowest excited state that has significant overlap with the vacuum is $n=4$. Because of this, this state continues to have the highest spectral coefficient, and therefore it dominates the correlator at short distances. On the other hand, the asymptotic long-distance behaviour is given by the lowest state $\Lambda_1$. Because the corresponding spectral coefficient is very small, it only starts to dominate the correlator at extremely long super-horizon distances. 

{When the potential $V(\phi)$ has two non-degenerate minima separated by a high barrier, the lowest state $n=1$ is localised in the false vacuum state, and therefore we can interpret it as the contribution from the domains of false vacuum which are occasionally formed. The eigenvalue $\Lambda_1$ represents the size of these domains, and the spectral coefficient is suppressed because of their rarity.}

{However, in the example shown in Fig.~\ref{fig:b6} this behaviour occurs even though the potential $V(\phi)$ has only one minimum and there is therefore no false vacuum state (see Fig.~\ref{fig:po1}).
In that case the lowest state $n=1$ corresponds to excursions of the field into high field values far away from the minimum.
}
\begin{figure}[t]\begin{center}
\includegraphics[width=0.80\textwidth,trim={0cm 0cm 0cm 7cm},clip]{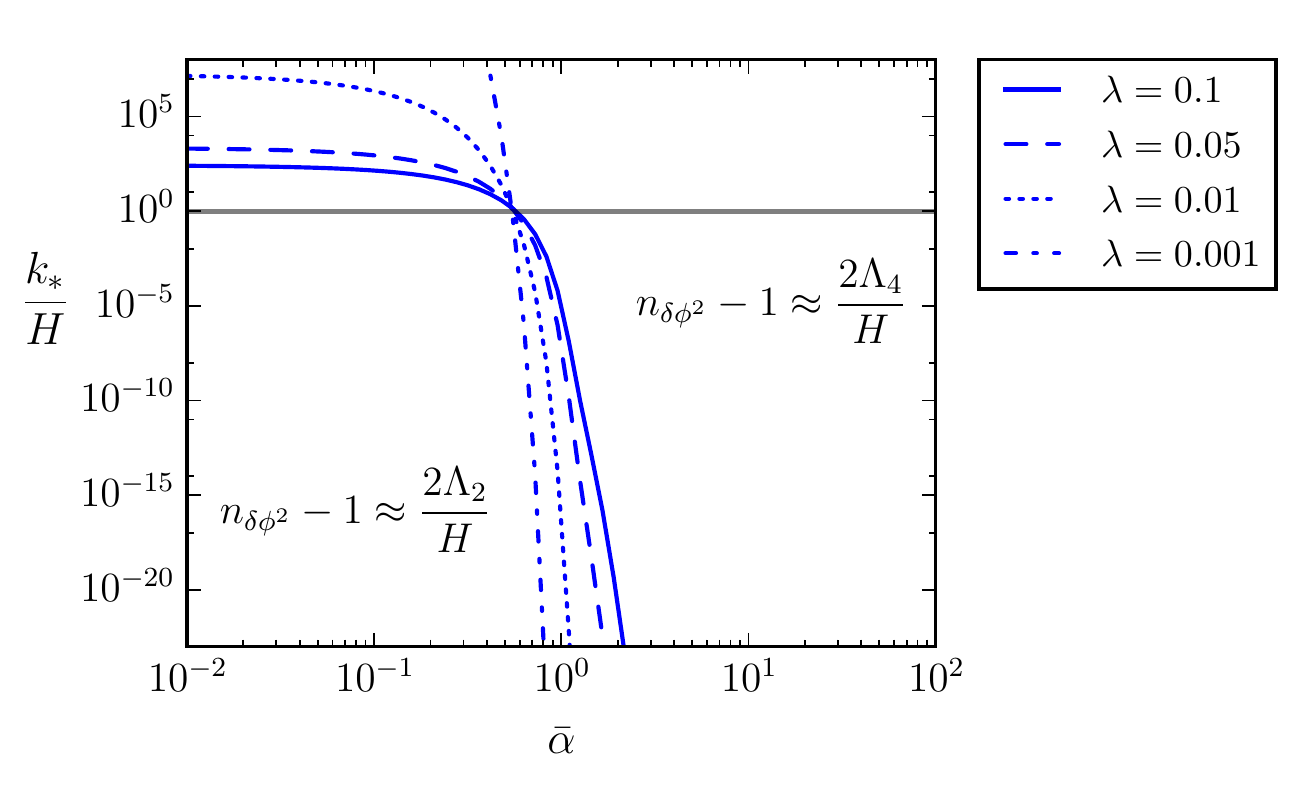}
\end{center}
\caption{The wavenumber $k_*$ at which the ${\cal P}^{(2)}_{\delta\phi^2}(k)$ term starts dominating the power spectrum (\ref{eq:spectrumfull}) over ${\cal P}^{(4)}_{\delta\phi^2}(k)$. Above the gray line the distances are sub-horizon and hence not amendable to a stochastic treatment. \label{fig:p2p4}
}
\end{figure}

As in the asymmetric case, it is possible that $\Lambda_1$ is so small that the false vacuum domains are larger than $1/H_0$. Then, because the true and false vacuum have different physical properties, the observables we would measure would depend on the vacuum we are in. This would not affect the eigenvalues in the spectral expansion (\ref{equ:spectralx}), but the spectral coefficients $f_n$ would have to be computed using the false vacuum ground state $\psi_1$ rather than the true ground state $\psi_0$.

\subsection{Power spectrum}
For cosmological observations, the power spectrum is often a more relevant quantity than the coordinate-space correlator.
Therefore it is useful to calculate the power spectrum by taking the Fourier transform of (\ref{equ:spectralx}) 
\be
{\cal P}_f(k)=\sum_n \f{2}{\pi}f^2_n\,\Gamma\bigg(2-2\f{\Lambda_n}{H}\bigg)\sin\bigg(\f{\Lambda_n\pi}{H}\bigg)\bigg(\f{k}{H}\bigg)^{2\Lambda_n/H}\equiv \sum_{n=1}{\cal P}^{(n)}_{f}(k)\,,
\label{eq:spectrumfull}
\ee
where $k$ is the physical momentum.

Assuming that a single $n=d$ term dominates the expansion, the spectral index $n_f$ can then be written in a simple form
\be
{n_{f}-1}\equiv\f{\ln{\cal P}_{f}(k)}{\ln k}\approx\f{\ln{\cal P}^{(d)}_{f}(k)}{\ln k}=\f{2\Lambda_d}{H}.\label{eq:SI}
\ee

When there are non-trivial hierarchies between the spectral coefficients, the spectral index $n_f$ can be different on different scales.
As an example, consider the power spectrum of $\delta\phi^2\equiv \phi^2-\langle\phi^2\rangle$ in the symmetric case. 
At asymptotically small wave number $k$, the spectral index is given by the lowest state $n=2$, but because its spectral coefficient is very small, the higher state $n=4$ dominates at higher $k$.
Comparing these two terms, we can straightforwardly solve for the wavenumber $k_*$ at which they cross as
\be
\f{{\cal P}^{(2)}_{\delta\phi^2}(k_*)}{{\cal P}^{(4)}_{\delta\phi^2}(k_*)}=1\quad\Leftrightarrow\quad\f{k_*}{H}=\Bigg\{\f{\left[(\phi^2)_2\right]^2\Gamma\big(2-2\f{\Lambda_2}{H}\big)\sin\big(\f{\Lambda_2\pi}{H}\big)}{\left[(\phi^2)_4\right]^2\,\Gamma\big(2-2\f{\Lambda_4}{H}\big)\,\sin\big(\f{\Lambda_4\pi}{H}\big)}\Bigg\}^{\f{H}{2(\Lambda_4-\Lambda_2)}}\,,
\ee
which is plotted in Fig.\,\ref{fig:p2p4}.
When $\bar{\alpha}\lesssim 1$, $k_*/H\gtrsim 1$, and therefore the first term $n=2$ dominates on all scales and the spectral index is therefore constant to a good approximation. However, when $\bar{\alpha}\gtrsim 1$, we can see that $k_*/H$ rapidly becomes very small, which means the crossover happens at scales that were massively superhorizon during inflation.
What this shows is that the scale at which the spectral index changes can for some parameter values occur at scales that are visible in the cosmic microwave background or other cosmological observations, possibly providing an important observational signature of early Universe models involving decoupled spectator scalars.

\section{Conclusions}
\label{sec:Conclusions}
In this work we have studied spectator scalar fields in de Sitter space with a potential of the double-well form (\ref{eq:pot}) by means of the stochastic spectral expansion \cite{Starobinsky:1994bd}. {The terms in the expansion are determined by eigenvalues and eigenfunctions which we solve numerically.} This work is a continuation of Ref.\,\cite{Markkanen:2019kpv} where only potentials with manifest $\mathbb{Z}_2$ symmetry and a single minimum were considered. 
{Our calculations show that also asymmetric} potentials with multiple minima can be efficiently studied with simple numerical methods to high precision, implying the technique to be rather powerful and suitable for a large class of potentials. In this vein we note two interesting possibilities that are yet unexplored: models with more than one scalar and/or potentials with periodic boundary conditions such as for the axion.

The double-well potential has unsurprisingly a much richer structure in terms of eigenfunctions and -values than the quartic or quadratic cases. 
{In particular, the spectrum has states that are not localised near the minimum of the potential and which are therefore not visible in perturbation theory. In most cases these non-perturbative states dominate the correlation function especially at long distances. The spectral coefficients of the terms can also be very different, which can lead to non-trivial behaviour such as a change of the spectral index at very long distances. If this happens at cosmologically relevant scales, it can be observable. Interestingly, this behaviour persists even when the potential has only one minimum.
In the symmetric case, the behaviour of odd quantities (such as the field) is also very different from the behaviour of even quantities (such as energy density). Together, these effects demonstrate that the naive intuition based perturbation theory is, in general, not applicable.}

Based on our analysis we conclude that spectator fields with a double-well potential have many possibly observable and currently unexplored cosmological consequences. 

\section*{Acknowledgments}
We thank Jacopo Fumagalli, Gabriel Moreau,  Julien Serreau and Sébastien Renaux-Petel for illuminating discussions. AR is supported by the U.K. Science and Technology Facilities Council grant ST/P000762/1 and an IPPP Associateship. 
This project has received funding
from the European Union’s Horizon 2020
research and innovation programme under the Marie Skłodowska-Curie grant agreement No. 786564.

\appendix
\section{Eigenfunctions and -values}
\label{sec:appendix}

\begin{figure}[t]\begin{center}
\includegraphics[width=1\textwidth,trim={0cm 0 0 0cm},clip]{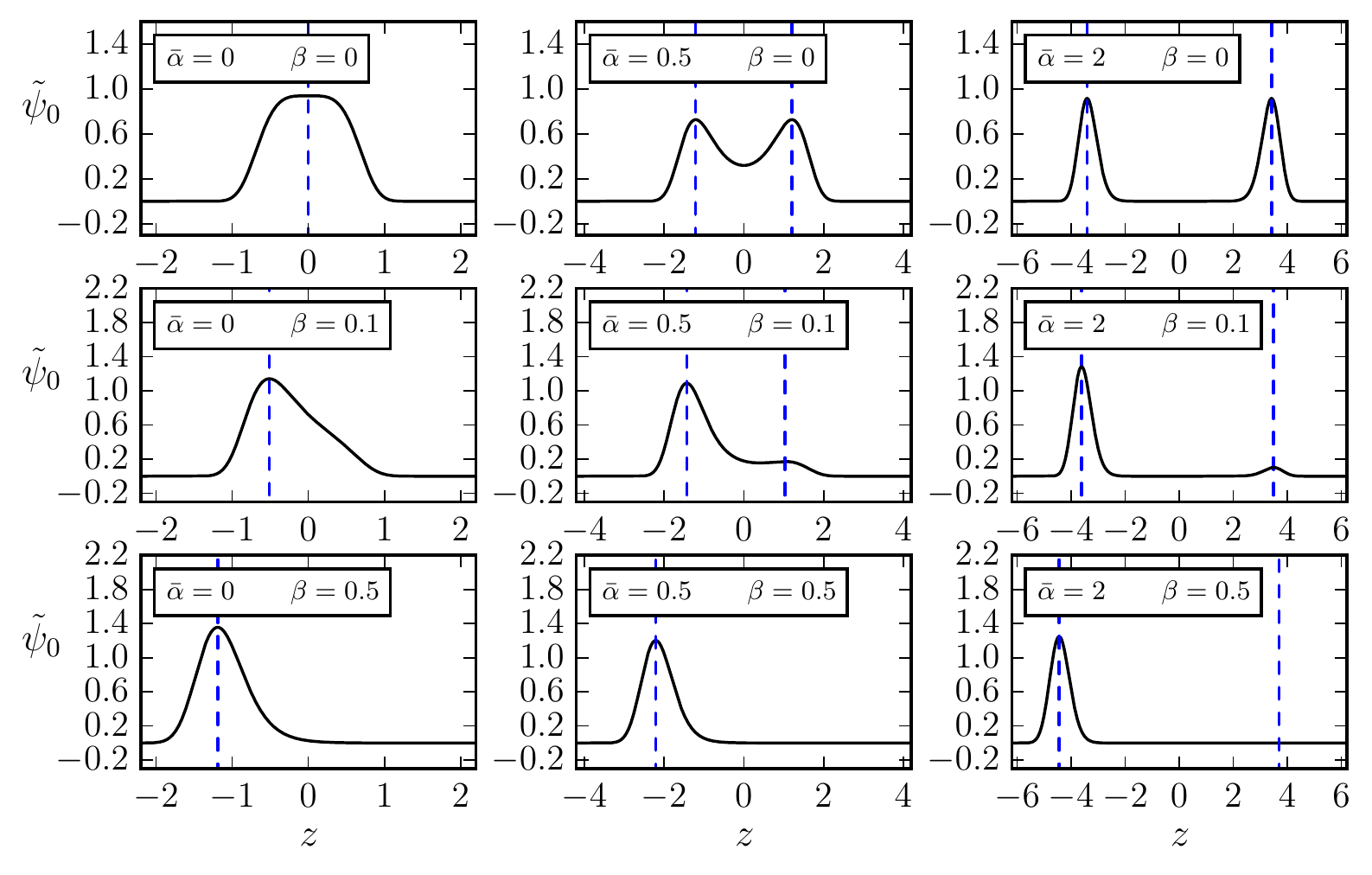}
\end{center}
\caption{
The dimensionless $n=0$ eigenfunctions from Eq.~(\ref{eq:quarz2}).
\label{fig:p1}}
\end{figure}

\begin{figure}[t]
\begin{center}
\includegraphics[width=1\textwidth,trim={0cm 0 0 0cm},clip]{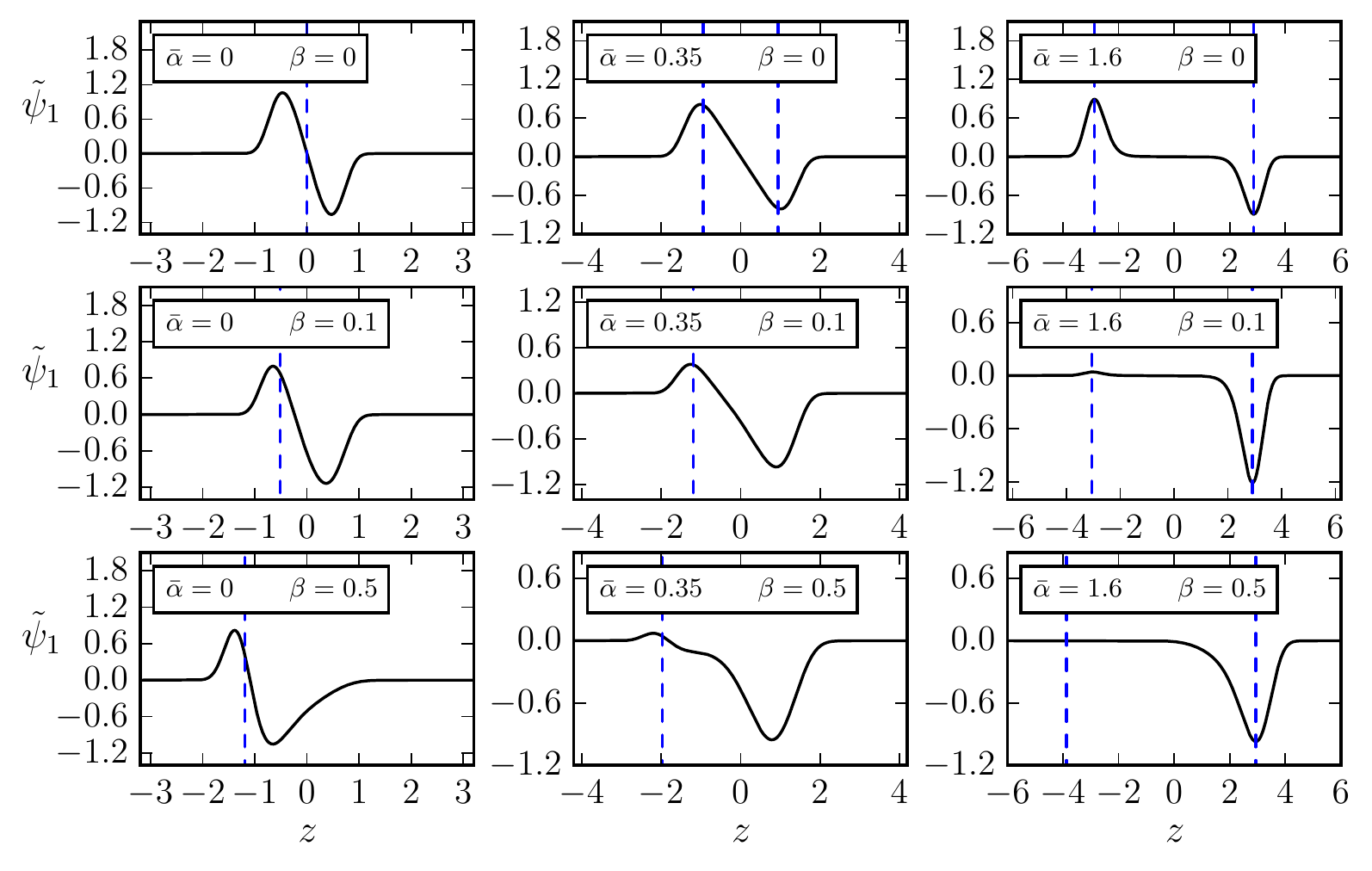}
\includegraphics[width=1\textwidth,trim={0cm 0 0 0cm},clip]{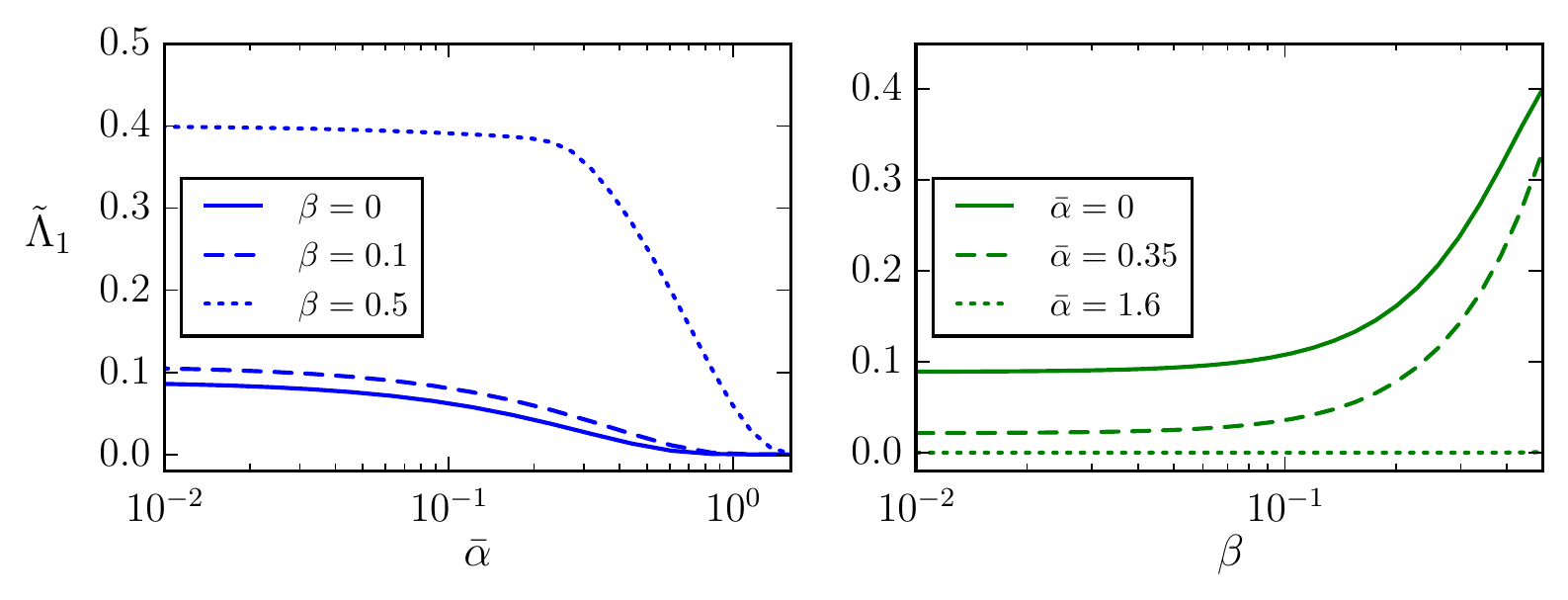}
\end{center}
\caption{
The dimensionless $n=1$ eigenfunctions and -values from Eq.~(\ref{eq:quarz2}).
\label{fig:p2}}
\end{figure}

\begin{figure}[t]
\begin{center}
\includegraphics[width=1\textwidth,trim={0cm 0 0 0cm},clip]{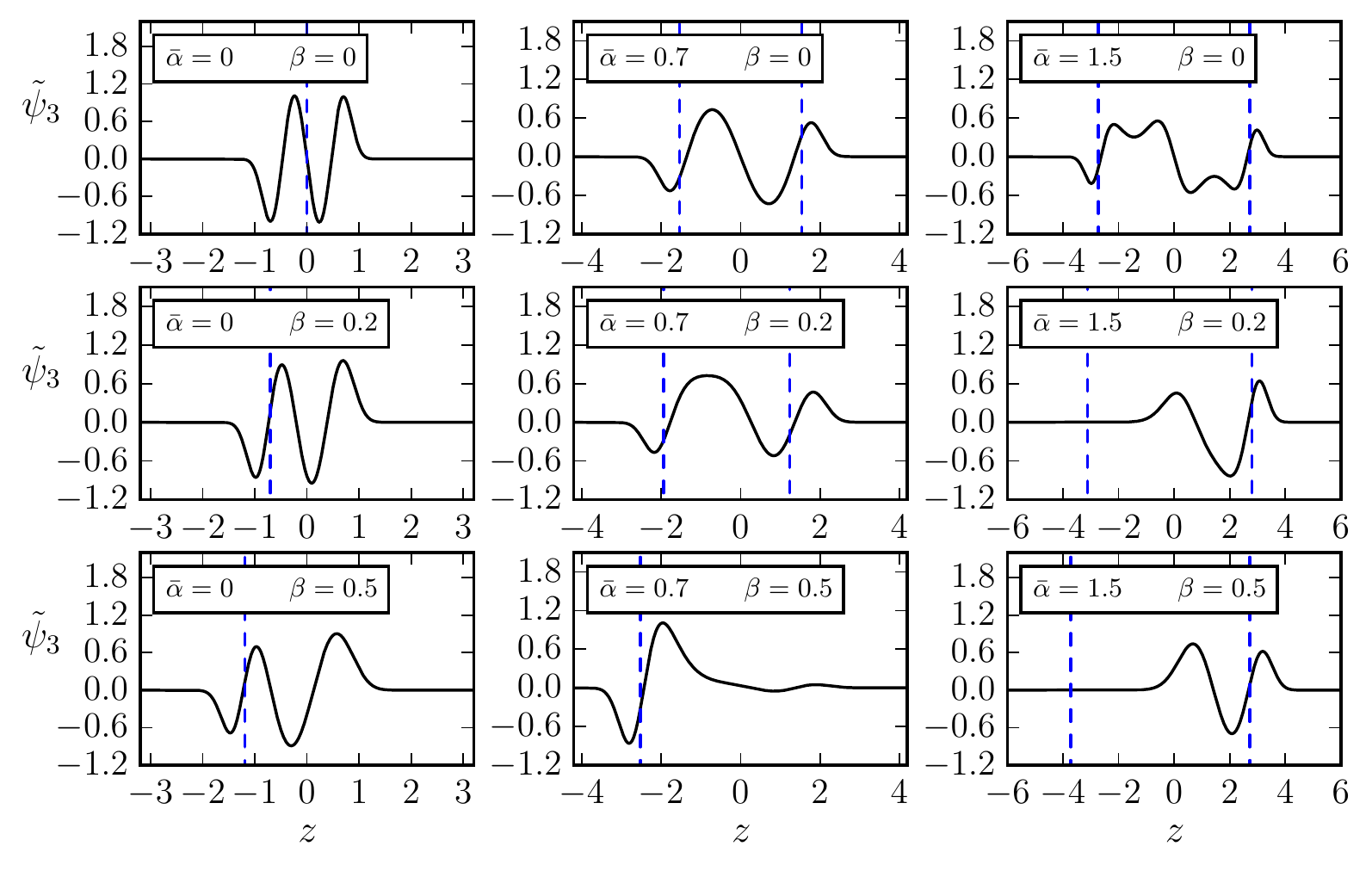}
\includegraphics[width=1\textwidth,trim={0cm 0 0 0cm},clip]{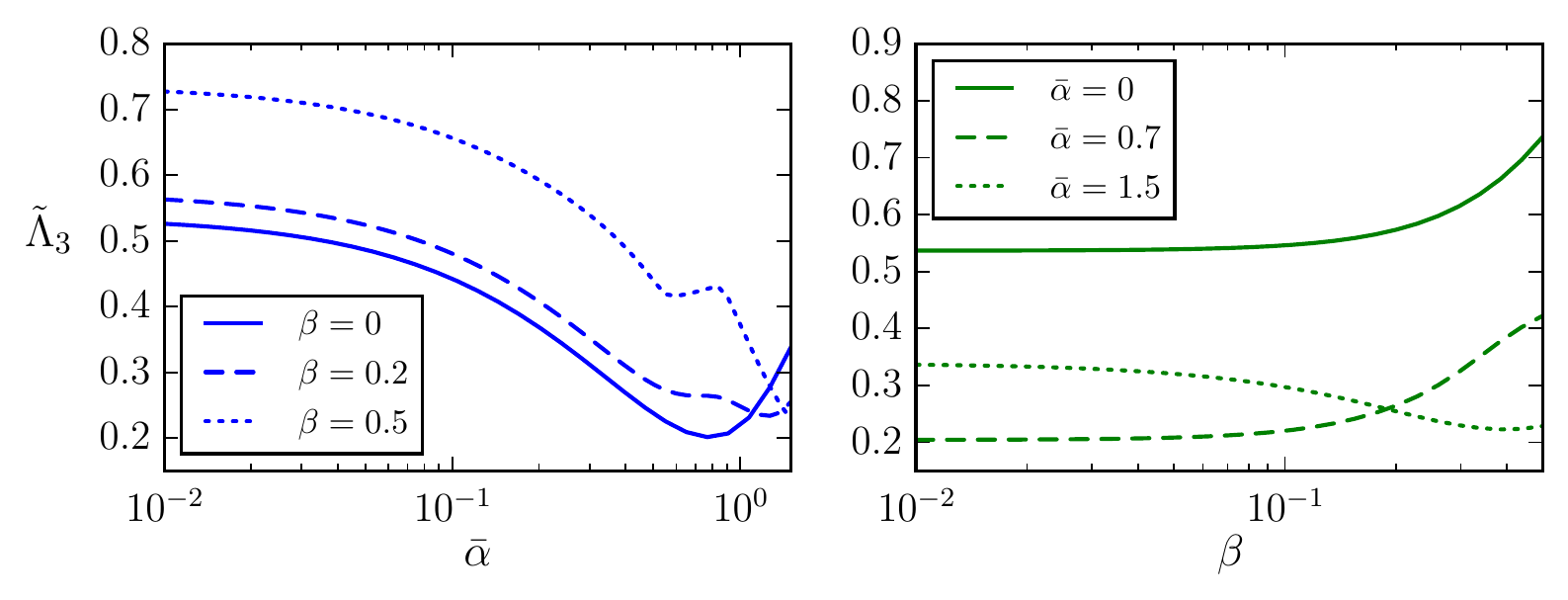}
\end{center}
\caption{
The dimensionless $n=3$ eigenfunctions and -values from Eq.~(\ref{eq:quarz2}).\label{fig:p4}}
\end{figure}

\begin{figure}[t]
\begin{center}
\includegraphics[width=1\textwidth,trim={0cm 0 0 0cm},clip]{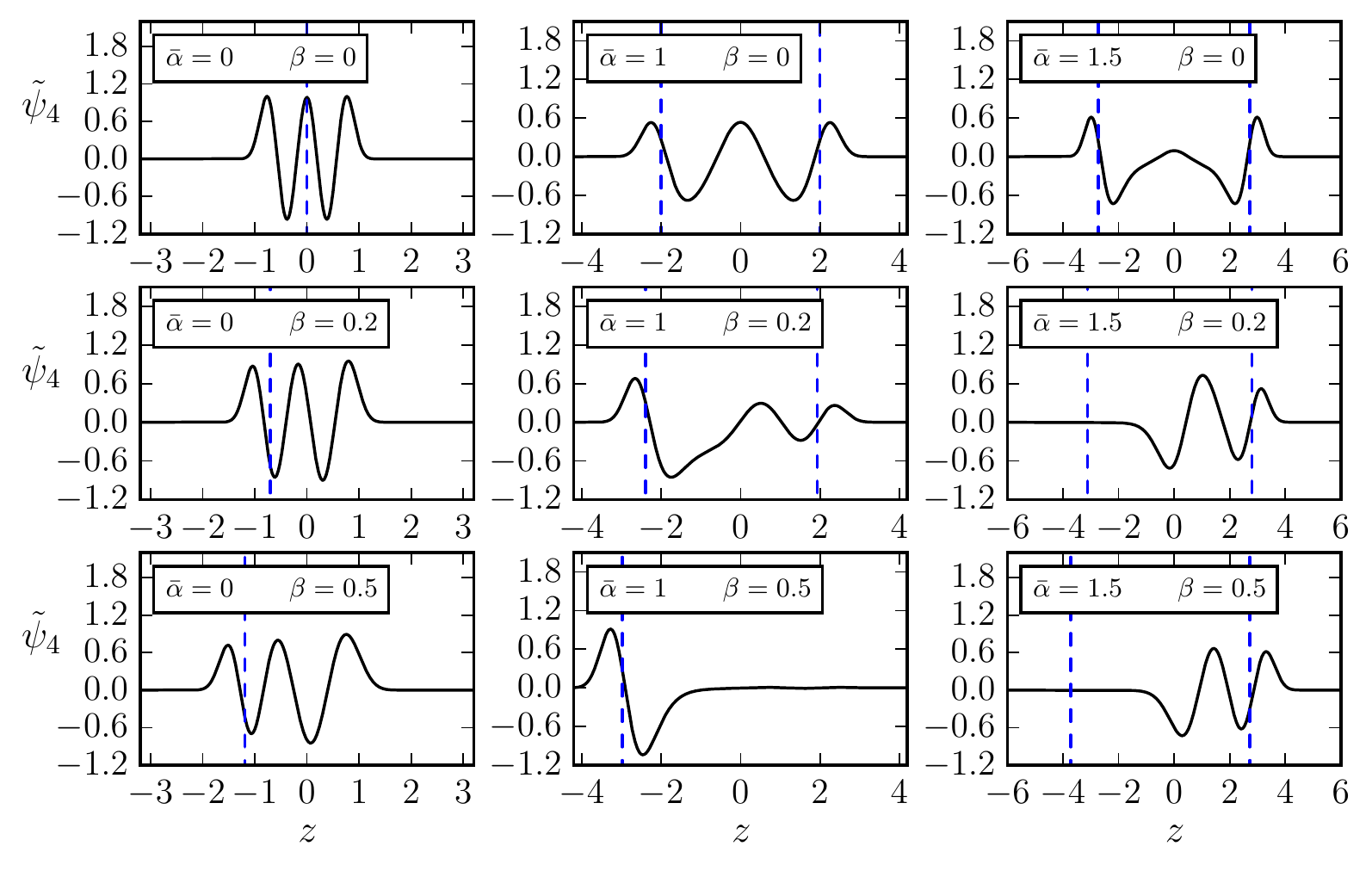}
\includegraphics[width=1\textwidth,trim={0cm 0 0 0cm},clip]{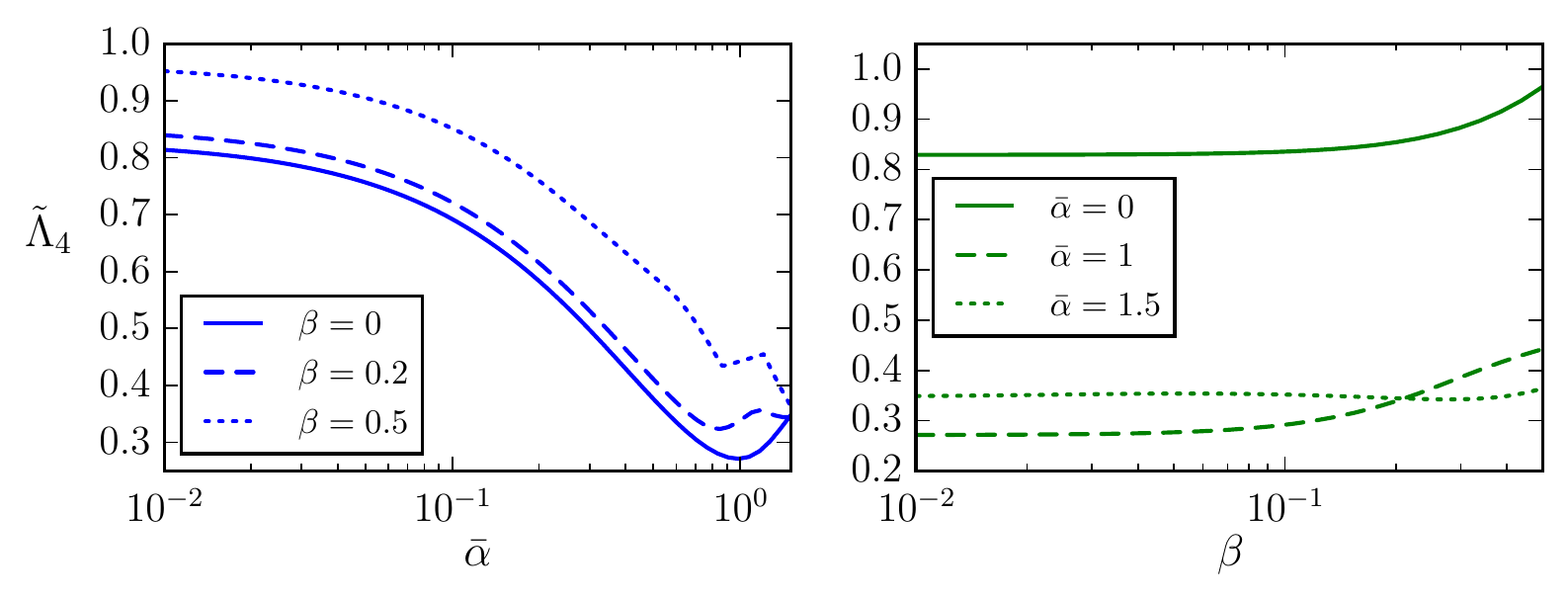}
\end{center}
\caption{
The dimensionless $n=4$ eigenfunctions and -values from -values from Eq.~(\ref{eq:quarz2}). \label{fig:p5}}
\end{figure}

{
Figs.\,\ref{fig:p1}--\ref{fig:p5} show the lowest eigenfunctions and eigenvalues for some representative choices for $\bar{\alpha}$ and $\beta$.
}
The various values for $\bar{\alpha}$ and $\beta$ have been specifically chosen as to in include as many qualitatively different cases as possible. The blue dashed lines indicate the locations of the minima. Note that in the non-degenerate case with $\beta\neq0$ the global minimum is located left of the origin. 

\begin{figure}[t]
\begin{center}
\includegraphics[width=1\textwidth,trim={0cm 0 0 0cm},clip]{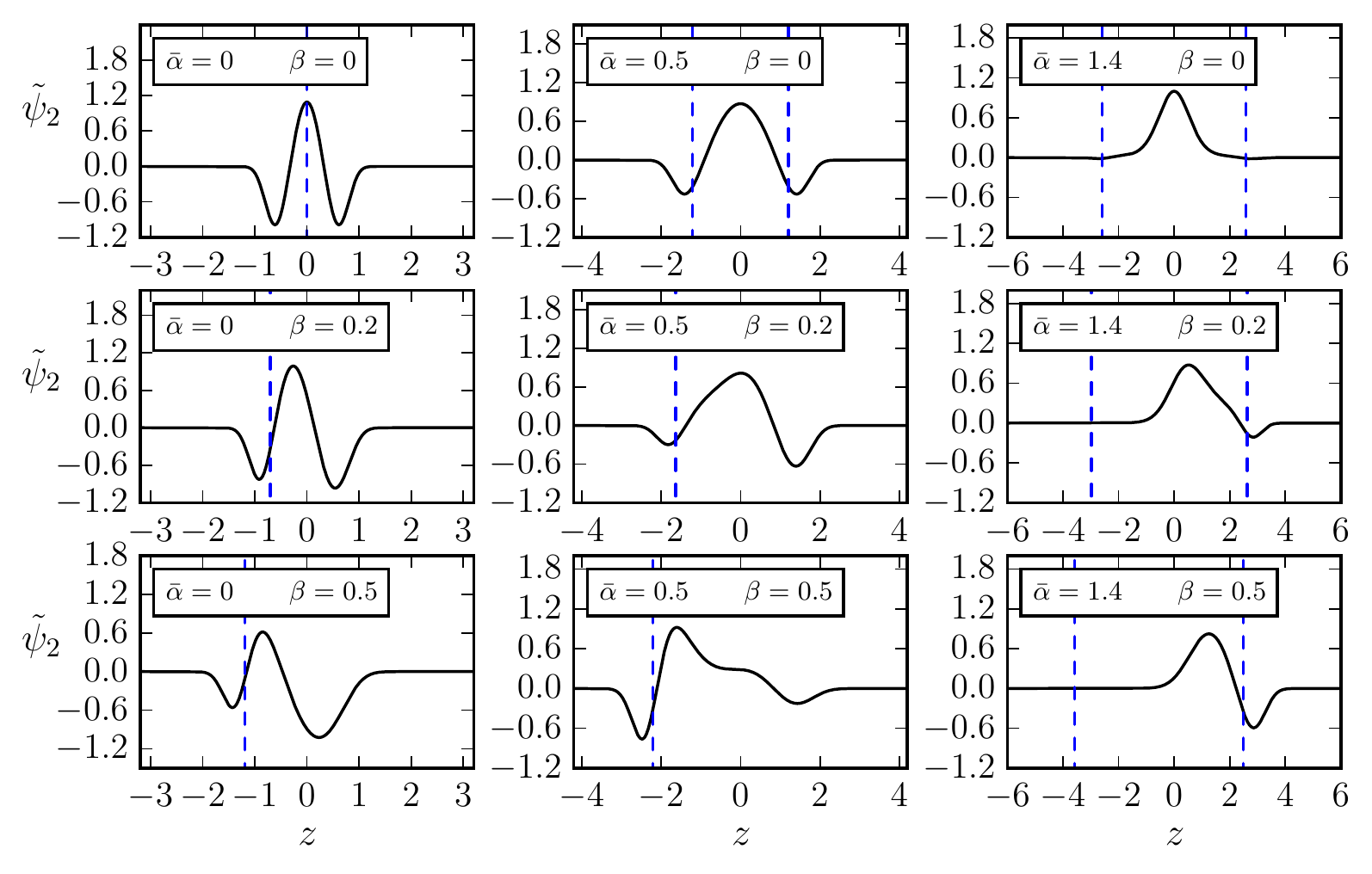}
\includegraphics[width=1\textwidth,trim={0cm 0 0 0cm},clip]{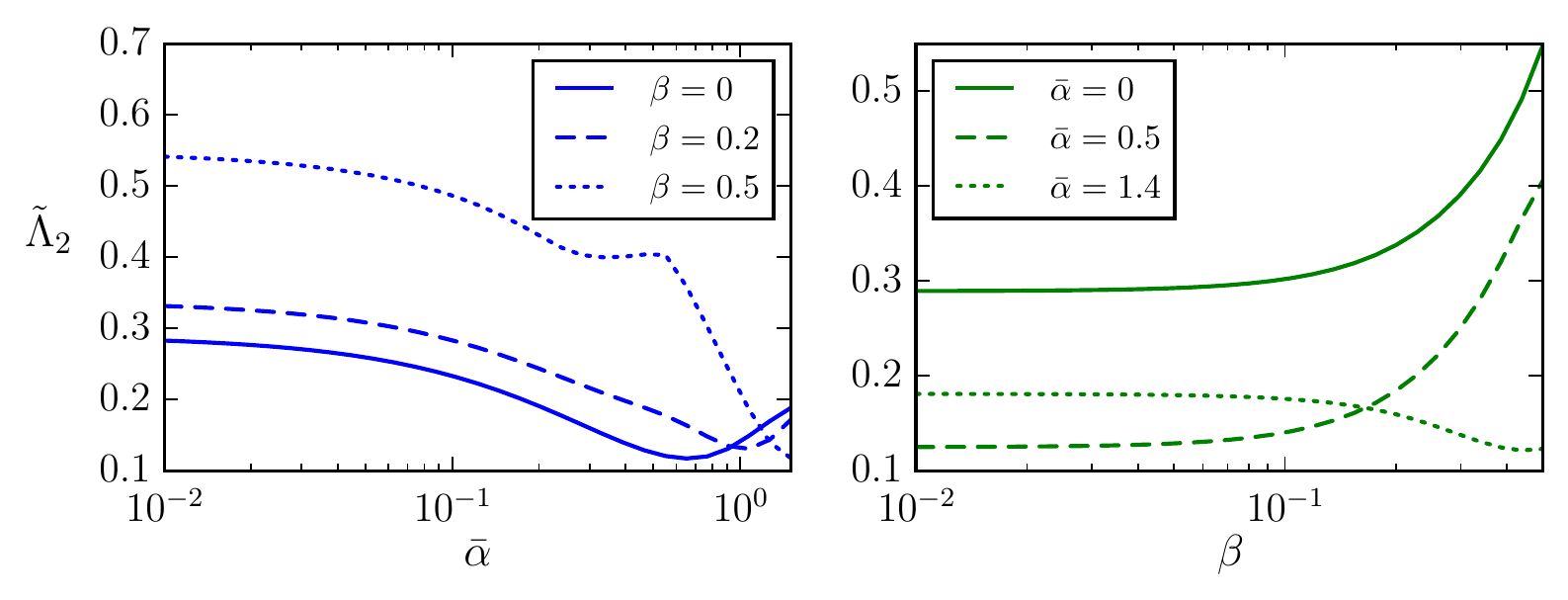}
\end{center}
\caption{
The dimensionless $n=2$ eigenfunctions and -values from Eq.~(\ref{eq:quarz2}).}
\end{figure}

\bibliography{HiggsFluctuations}


\end{document}